\documentclass[a4paper,11pt]{article}
\pdfoutput=1
\usepackage{jcappub}

\usepackage{amsmath}
\usepackage{amssymb}
\usepackage{graphicx}
\newcommand{\figpath}[1]{#1}

\title{Spin Polarization of Proca Stars Formed by Gravitational Bose--Einstein Condensation}

\author[a]{Jiajun Chen}
\affiliation[a]{School of Physical Science and Technology, Southwest University, Chongqing 400715, China}
\emailAdd{chenjiajun@swu.edu.cn}

\abstract{We study the internal spin polarization of Proca stars formed by gravitational Bose--Einstein condensation of a three-component nonrelativistic vector field.  In idealized periodic-box simulations, we decompose the aperture-averaged spin into a coherent net fraction, a local polarization fraction, and their ratio, thereby distinguishing genuine coherent core polarization from local spin density whose direction cancels inside the aperture.  For independent vector components, condensation produces Proca stars that are sizably but not maximally polarized.  Across an independent-component simulation ensemble, the coherent core-spin fraction has mean $\langle\chi_{\rm net}\rangle\simeq0.62$, with substantial realization-to-realization scatter.  We interpret this scatter as the outcome of random elliptical polarization of the dominant component-space mode, rather than as evidence for a universal Proca-star spin fraction.  This interpretation is supported by the core polarization matrix: its leading eigenvector provides an estimate of the ideal single-mode spin fraction, while the difference between this estimate and the directly integrated coherent spin tracks the departure of the core from a rank-one component-space state.  The measured leading-eigenvector spin fractions are broadly compatible with an isotropic random-complex-vector model and less compatible with an equal-amplitude random-phase model.  Correlated and circular initial data drive the dominant component-space mode toward the circular-polarization bound, giving the ordering independent $\rightarrow$ correlated $\rightarrow$ circular.  These results show that internal polarization is a genuine vector degree of freedom of gravitationally condensed nonrelativistic Proca stars, and that the resulting core spin is controlled by the polarization of the dominant condensed mode rather than by a fixed universal value.}

\keywords{vector dark matter, Proca stars, gravitational Bose--Einstein condensation, dark matter simulations, spin polarization}

\compress

\begin{document}

\maketitle

\section{Introduction}
\label{sec:introduction}

Ultralight bosonic dark matter provides a wave-mechanical alternative to collisionless cold dark matter on sufficiently small scales \cite{Aghanim:2018eyx,hu2000,Sikivie:2006ni,Arvanitaki:2009fg,Hui:2016ltb,Marsh:2015xka,Suarez:2013iw,Hlozek:2017zzf}.  In scalar fuzzy dark matter, gravitational relaxation produces compact solitonic cores whose structure is described mainly by density and phase \cite{Schive:2014dra,Schive:2014hza,Veltmaat:2018dfz,Bar:2018acw,Du:2016aik,Mocz:2017wlg}.  Vector dark matter, including hidden photon dark matter, contains additional spin and polarization degrees of freedom \cite{Graham:2015rva,PhysRevD.108.083021,PhysRevD.111.043031}.  Polarized spin-1 and higher-spin solitons can carry macroscopic intrinsic spin and admit partially polarized states~\cite{Jain:2021pnk}.  Here ``Proca star'' denotes the nonrelativistic self-gravitating vector soliton, distinguished from relativistic complex-field Einstein--Proca stars~\cite{Amin:2022pkn,Jain:2021pnk,Gorghetto:2022sue,Volkov:2002aj,Zhang:2024bjo}.

The formation of compact objects by gravitational Bose--Einstein condensation has been studied extensively for self-gravitating bosonic fields \cite{1991PhRvL..66.1659S,PhysRevD.42.384,1994PhRvL..72.2516S,Liddle:1993ha,Kolb:1993zz,Kolb:1993hw,PhysRevD.84.043531,Chavanis:2011zm,Chavanis:2016dab,Eby:2015hsq,Levkov:2018kau,PhysRevD.104.083022,PhysRevD.106.023009,Amin:2019ums,Kirkpatrick:2020fwd,Eggemeier:2019jsu,Jain:2023ojg,Jain:2024krs}.  Related Schr\"odinger-Poisson simulations have explored soliton formation, mergers, core-halo relations, vortex dynamics, and wave interference \cite{Widrow:1993qq,Schwabe_2016,Mocz:2015sda,Uhlemann:2014npa,Hui_2021,Helfer_2019,Chen_2024,Zeng:2026vortex}.  Multicomponent Schr\"odinger-Poisson condensation, including spin-$s$ systems, has been studied in the context of kinetic relaxation and Bose-star formation~\cite{Jain:2023ojg}.  For vector fields, previous work showed that condensation depends on correlations among vector components and that the growing Proca star can acquire spin angular momentum \cite{PhysRevD.108.083021}.  Vector-dark-matter halo studies have also found intrinsically and partially polarized solitonic cores, including cases with small total halo spin~\cite{Amin:2022pkn,PhysRevD.111.043031}.  This observation raises a core-level question: when independent complex vector components condense into the same gravitationally bound density peak, does the Proca star acquire a coherent net spin, or do locally polarized regions cancel directionally inside the core?

Answering this requires diagnostics that tell local polarization apart from coherent aperture-averaged spin.  Gravity couples to the total density, but the spin density depends on the relative amplitudes and phases of the vector components, so a large local $|\mathbf{s}|$ need not add up to a coherent core spin.  We therefore decompose the aperture spin into a net fraction $\chi_{\rm net}$, a local fraction $\chi_{\rm loc}$, and the ratio $\eta=\chi_{\rm net}/\chi_{\rm loc}$, and we use the core polarization matrix to connect the measured spin to the component-space state of the condensed mode.

In our independent realizations $\chi_{\rm net}$ is sizable and $\eta$ sits near unity, so the aperture-averaged spin is coherent rather than a local spin fluctuation.  Its value is seed dependent, and both the typical size and the realization-to-realization scatter are what one expects if the Proca star is dominated by a single complex component-space vector whose elliptical polarization is drawn at random.

We test the dominant-mode interpretation using the core polarization matrix.  Its leading eigenvector predicts the spin of the dominant component-space mode, and the small gap to the directly integrated aperture spin tracks the departure from a rank-one core, vanishing as the core becomes exactly single-mode.  The core spin fraction is therefore not a fixed number but one set by the polarization of the dominant mode, and the earlier core spin per particle of order $|S_c|/N_c\sim0.8$ measured in the same idealized setting~\cite{PhysRevD.108.083021} is naturally interpreted as a high-spin realization within a broad random-polarization outcome.

This paper is organized as follows.  In Sec.~\ref{sec:eom} we summarize the nonrelativistic vector Schr\"odinger--Poisson system and define the aperture spin fractions.  In Sec.~\ref{sec:simulations} we introduce the dimensionless initial conditions and the independent, correlated, and circular polarization preparations.  In Sec.~\ref{sec:results} we present the core-spin results and their component-space interpretation.  We conclude in Sec.~\ref{sec:conclusion}.

\section{Schr\"odinger--Poisson System and Core Spin}
\label{sec:eom}

\subsection{Equations and Spin Fractions}
\label{subsec:sp-spin-diagnostics}

A real vector field $A_\mu$, with mass $m$, minimally coupled to gravity, has the action
\begin{equation}
S =\int d^4 x \sqrt{-g}\left( -\frac{1}{4}F_{\mu\nu}F^{\mu\nu}-\frac{1}{2}m^2 A_\mu A^\mu\right),
\label{eq:action}
\end{equation}
where $F_{\mu\nu}=\partial_\mu A_\nu-\partial_\nu A_\mu$ is the field-strength tensor, and indices are raised and lowered by the metric $g_{\mu\nu}$ with determinant $g$.  The equation of motion is
\begin{equation}
\nabla_\mu F^{\mu\nu}= m^2 A^\nu,
\label{eq:proca_eq}
\end{equation}
known as the Proca equation.  It implies the Lorenz constraint $\nabla_\mu A^\mu=0$, reducing the number of physical degrees of freedom of $A_\mu$ to three dynamical fields (see e.g. Refs.~\cite{Brito:2015pxa,Gorghetto:2022sue}).  In the nonrelativistic limit, these degrees of freedom are written as
\begin{equation}
A_i=\frac{1}{\sqrt{2m}}\left(\psi_i e^{imt}+\psi_i^\ast e^{-imt}\right).
\label{eq:A_i}
\end{equation}
In exact analogy with a real scalar field (see e.g. Ref.~\cite{Marsh:2015xka}), for small perturbations of the metric about Minkowski space by the Newtonian potential $\Phi$, and at lowest order in the nonrelativistic limit, the field equations reduce to three copies of the Schr\"odinger-Poisson equations for the complex fields $\psi_i$ \cite{Jain:2021pnk,Jain:2023ojg}, \cite{Mendonca:2021aeq,Salehian:2021khb,PhysRevD.108.083021,PhysRevD.111.043031},
\begin{equation}
i\frac{\partial}{\partial t}\boldsymbol{\psi}
=-\frac{1}{2m}\nabla^2\boldsymbol{\psi}
+m\Phi\boldsymbol{\psi},
\label{eq:SP1}
\end{equation}
\begin{equation}
\nabla^2\Phi=4\pi Gm\left(\boldsymbol{\psi}^{\dagger}\boldsymbol{\psi}-\bar n\right),
\label{eq:SP2}
\end{equation}
where
\begin{equation}
\boldsymbol{\psi}=(\psi_x,\psi_y,\psi_z).
\end{equation}
Here $\boldsymbol{\psi}^{\dagger}\boldsymbol{\psi}$ is the total number density, $\bar n$ is its spatial mean, $G$ is Newton's gravitational constant, and natural units, $\hbar=c=1$, are used.

Introducing a reference velocity $v_0$, the standard dimensionless variables are \cite{PhysRevD.106.023009,PhysRevD.108.083021}
\begin{equation}
\begin{gathered}
\widetilde{\mathbf{x}}=m v_0\mathbf{x},\qquad
\widetilde{t}=m v_0^2 t,\qquad
\widetilde{\Phi}=\frac{\Phi}{v_0^2},
\\
\widetilde{\boldsymbol{\psi}}
=\left(\frac{4\pi G}{m v_0^4}\right)^{1/2}\boldsymbol{\psi},
\qquad
\widetilde{\bar n}
=\frac{4\pi G}{m v_0^4}\bar n.
\end{gathered}
\end{equation}
The nonrelativistic equations become
\begin{equation}
i\partial_{\widetilde{t}}\widetilde{\boldsymbol{\psi}}
=-\frac{1}{2}\widetilde{\nabla}^{2}\widetilde{\boldsymbol{\psi}}
+\widetilde{\Phi}\widetilde{\boldsymbol{\psi}},
\end{equation}
with
\begin{equation}
\widetilde{\nabla}^{2}\widetilde{\Phi}
=\widetilde{\boldsymbol{\psi}}^{\dagger}\widetilde{\boldsymbol{\psi}}
-\widetilde{\bar n}.
\end{equation}

The density is
\begin{equation}
\rho=\boldsymbol{\psi}^{\dagger}\boldsymbol{\psi}
=|\psi_x|^2+|\psi_y|^2+|\psi_z|^2.
\end{equation}
Gravity therefore couples only to the total density, while the internal vector nature of the condensed object is carried by the relative amplitudes and phases of the three components.  The corresponding spin density is
\begin{equation}
\mathbf{s}=i\boldsymbol{\psi}\times\boldsymbol{\psi}^{\ast}.
\label{eq:spin-density}
\end{equation}

For a spherical aperture of radius $R$ centered on the density maximum, we use the volume averages
\begin{equation}
\begin{aligned}
V_R&=\int_{r<R}d^3x, &
\langle\rho\rangle_R&=\frac{1}{V_R}\int_{r<R}\rho\,d^3x,\\
\langle\mathbf{s}\rangle_R&=\frac{1}{V_R}\int_{r<R}\mathbf{s}\,d^3x, &
\langle|\mathbf{s}|\rangle_R&=\frac{1}{V_R}\int_{r<R}|\mathbf{s}|\,d^3x .
\end{aligned}
\end{equation}
These quantities describe the average density and spin density inside the aperture.  The corresponding integrated particle number and spin are
\begin{equation}
N_c(R)=\int_{r<R}\rho\,d^3x,\qquad
\mathbf{S}_c(R)=\int_{r<R}\mathbf{s}\,d^3x,
\end{equation}
and the three dimensionless spin diagnostics are
\begin{equation}
\chi_{\rm net}(R)=\frac{|\mathbf{S}_c(R)|}{N_c(R)},\qquad
\chi_{\rm loc}(R)=\frac{\int_{r<R}|\mathbf{s}|\,d^3x}{N_c(R)},\qquad
\eta(R)=\frac{\chi_{\rm net}(R)}{\chi_{\rm loc}(R)} .
\end{equation}
Here $\chi_{\rm net}$ measures the coherent spin per particle, $\chi_{\rm loc}$ measures the amount of local polarization, and $\eta$ measures directional coherence.  A large $\chi_{\rm loc}$ with small $\eta$ would indicate locally polarized regions whose spin directions cancel in the aperture average.

Writing $\boldsymbol{\psi}=\mathbf{a}+i\mathbf{b}$ gives
\begin{equation}
\rho=|\mathbf{a}|^2+|\mathbf{b}|^2,\qquad
\mathbf{s}=2\mathbf{a}\times\mathbf{b},\qquad
\frac{|\mathbf{s}|}{\rho}\leq 1 .
\end{equation}
The upper limit corresponds to a fully circularly polarized state with $|\mathbf{a}|=|\mathbf{b}|$ and $\mathbf{a}\perp\mathbf{b}$, while the lower endpoint has vanishing local spin.  Integrating this pointwise bound over the aperture likewise gives $\chi_{\rm loc}(R)\leq1$, with equality only if the core is pointwise circularly polarized throughout $r<R$.

We define the reference core aperture from the spherically averaged shell profile $\bar\rho_{\rm sh}(r)$ around the density maximum.  The core radius $R_c$ is the first radius at which $\bar\rho_{\rm sh}(r)=\rho_{\rm max}/2$, following the standard half-density convention for wave-dark-matter cores~\cite{Schive:2014dra}.  This shell-averaged radius is distinct from the volume averages used in the spin diagnostics.

\subsection{Polarization Matrix and the Single-Mode Limit}
\label{subsec:core-polarization-matrix}

The aperture-integrated spin fractions tell us how much coherent spin is present, while the component structure of that spin is encoded in the normalized core polarization matrix
\begin{equation}
P_{ij}(R)=\frac{1}{N_c(R)}\int_{r<R}\psi_i^\ast(\mathbf{x})\psi_j(\mathbf{x})\,d^3x.
\label{eq:polarization-matrix}
\end{equation}
This Hermitian matrix has unit trace.  Its imaginary antisymmetric part is not an additional observable: it is exactly the coherent spin per particle,
\begin{equation}
\frac{\mathbf{S}_c(R)}{N_c(R)}
=2\left({\rm Im}\,P_{yz},\,{\rm Im}\,P_{zx},\,{\rm Im}\,P_{xy}\right),
\label{eq:spin-from-p}
\end{equation}
and therefore
\begin{equation}
\chi_{\rm net}(R)
=2\left[({\rm Im}\,P_{xy})^2+({\rm Im}\,P_{yz})^2+({\rm Im}\,P_{zx})^2\right]^{1/2}.
\label{eq:chi-from-p}
\end{equation}
The real symmetric part fixes the polarization axes.  The quantities $\lambda_1,\lambda_2,\lambda_3$ are obtained by diagonalizing $P$,
\begin{equation}
\sum_j P_{ij}\,e_{a,j}=\lambda_a e_{a,i},\qquad a=1,2,3,
\label{eq:p-eigenproblem}
\end{equation}
where $\mathbf e_a$ are orthonormal component-space eigenvectors.  Since $P$ is Hermitian and positive semidefinite, the $\lambda_a$ are real and non-negative.  We order them as $\lambda_1\geq\lambda_2\geq\lambda_3$ and denote the leading eigenvector by $\mathbf c_1\equiv\mathbf e_1$.  Since $\mathrm{Tr}\,P=1$,
\begin{equation}
\lambda_1+\lambda_2+\lambda_3=1.
\end{equation}
The purity
\begin{equation}
\mathcal P=\mathrm{Tr}\,P^2=\lambda_1^2+\lambda_2^2+\lambda_3^2
\label{eq:purity}
\end{equation}
measures how close the core is to a single component-space direction.  The spin fraction carried by the leading eigenvector is
\begin{equation}
\chi_{\rm eig}=\frac{|i\mathbf{c}_1\times\mathbf{c}_1^\ast|}{\mathbf{c}_1^\dagger\mathbf{c}_1}.
\label{eq:chi-eig}
\end{equation}
For the normalized eigenvectors used below the denominator is unity; it is retained to make the expression independent of the chosen normalization of $\mathbf c_1$.
A useful reference point is the single-mode limit.  In this limit the three components share one spatial wave function $\phi(\mathbf{x},t)$, which describes the common core profile and phase evolution; this is not the gravitational potential $\Phi$.  Their relative amplitudes and phases are collected in a complex component-space vector $\mathbf c$ with entries $c_i$; the overall normalization of $\mathbf c$ is irrelevant in the ratios below.  If
\begin{equation}
\psi_i(\mathbf{x},t)\simeq c_i\,\phi(\mathbf{x},t),
\label{eq:single-mode-p}
\end{equation}
then
\begin{equation}
P_{ij}\simeq\frac{c_i^\ast c_j}{\mathbf{c}^\dagger\mathbf{c}},
\label{eq:p-single-mode}
\end{equation}
so $P$ has eigenvalues $(1,0,0)$ and $\mathcal P=1$.  Its leading eigenvector is $\mathbf{c}^{\ast}$ up to an irrelevant phase, and the spin fraction in Eq.~\eqref{eq:chi-eig} is invariant under this conjugation.  Substituting Eq.~\eqref{eq:single-mode-p} into the spin density gives $\mathbf s(\mathbf x,t)\simeq i(\mathbf c\times\mathbf c^\ast)\,|\phi(\mathbf x,t)|^2$ pointwise, so the common factor $|\phi|^2$ cancels identically between $\mathbf S_c(R)$ and $N_c(R)$ for every aperture $R$, giving $\chi_{\rm net}(R)\to|i\mathbf c\times\mathbf c^\ast|/(\mathbf c^\dagger\mathbf c)$, which is exactly $\chi_{\rm eig}$ by the conjugation invariance just noted.  In this limit $\chi_{\rm net}=\chi_{\rm eig}$ therefore holds as a strict, aperture-independent equality rather than an approximation.  Departures from this rank-one form quantify residual excited modes, imperfect spatial overlap among components, and finite-aperture mixing.

\section{Simulations and Initial Conditions}
\label{sec:simulations}

The dimensionless three-component Schr\"odinger-Poisson system is evolved in a periodic cubic domain with a noncosmological background.  The numerical evolution uses fourth-order pseudospectral methods.  The component label $\widetilde M_i$ denotes the dimensionless particle-number normalization,
\begin{equation}
  \int |\widetilde\psi_i|^2\,d^3\widetilde{x}=4\pi\widetilde M_i ,
\end{equation}
so it is a normalization label rather than a dimensional mass.  The explored dimensionless range has populated components with $\widetilde M_i\simeq40$--$120$ and box sizes $\widetilde L\simeq10$--$30$.

Following the initial-condition prescription used in Ref.~\cite{PhysRevD.108.083021}, each populated component is generated as a homogeneous and isotropic Gaussian random field in the periodic box.  The momentum distribution is concentrated on a Dirac-$\delta$ shell,
\begin{equation}
  |\widetilde\psi_i(\widetilde{\mathbf k})|^2
  \propto \delta(|\widetilde{\mathbf k}|-\widetilde k_0),
  \qquad \widetilde k_0=1,
\end{equation}
with random phases on the populated Fourier modes.  The resulting real-space fields are normalized to the desired component particle numbers.  This construction gives an initially homogeneous and isotropic field without an inserted compact core.

The three component-space preparations are:
\begin{itemize}
\item \textit{Independent components.}
\begin{equation}
\psi_x={\cal N}_x u_x,\qquad
\psi_y={\cal N}_y u_y,\qquad
\psi_z={\cal N}_z u_z.
\end{equation}
Here $u_x,u_y,u_z$ are independent complex random fields drawn on the $\delta$ shell, and ${\cal N}_i$ denotes component-wise normalization.  This preparation imposes no phase relation among the vector components.

\item \textit{Circular $xy$ preparation.}
\begin{equation}
\psi_x=\frac{u}{\sqrt{2}},\qquad
\psi_y=-\frac{i u}{\sqrt{2}},\qquad
\psi_z=0,
\label{eq:circular-xy-ic}
\end{equation}
where $u$ is one $\delta$-shell complex random field.  This state is pointwise circularly polarized in the $xy$ plane, with $|\mathbf{s}|/\rho=1$.
Because all components evolve with the same scalar Hamiltonian, the relation $\psi_y=-i\psi_x$ and $\psi_z=0$ is preserved by the equations of motion.  This case is therefore an analytic circular-polarization control as well as a numerical check.

\item \textit{Partially correlated components.}
\begin{equation}
\psi_x={\cal N}_x(\alpha u_x+\beta u_c),\qquad
\psi_y={\cal N}_y(\alpha u_y-i\beta u_c),\qquad
\psi_z={\cal N}_z u_z.
\end{equation}
Here $\alpha=\sqrt{1-f_{\rm corr}}$ and $\beta=\sqrt{f_{\rm corr}}$.  The field $u_c$ is a common $\delta$-shell random field shared by the $x$ and $y$ components, while $u_x$, $u_y$, and $u_z$ are independent.  We take $f_{\rm corr}=0.5$ and normalize each component to its prescribed particle number.  This preparation retains independent fluctuations and an independent $z$ component, but supplies a shared $xy$ channel that can be preferentially amplified by condensation and biases the core toward circular polarization.
\end{itemize}

The correlated and circular cases isolate the role of component correlations.  After late-time core formation, spin diagnostics are averaged over the condensed stage; uncertainties quoted for a single realization are temporal standard deviations over this stage.  The same diagnostics are applied to the independent, partially correlated, and circular preparations.  The independent-component statistics below use a broad set of simulations.

\section{Core Spin Results}
\label{sec:results}

\subsection{Aperture Spin of the Condensed Proca Star}

\begin{figure}[!htbp]
\centering
\includegraphics[width=0.98\linewidth]{\figpath{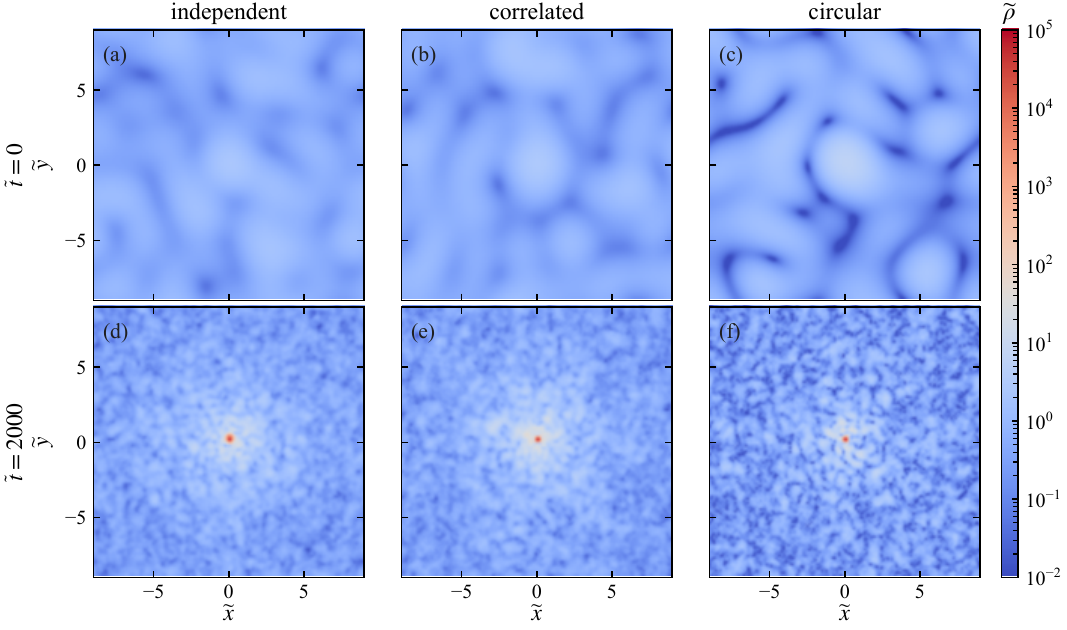}}
\caption{Snapshots of the dimensionless density field for three polarization preparations with $\widetilde{L}=18$.  The columns show independent, partially correlated, and circular $xy$ preparations with $(\widetilde{M}_x,\widetilde{M}_y,\widetilde{M}_z)=(80,80,80)$, $(80,80,80)$, and $(120,120,0)$, respectively.  Panels (a)--(c) show density slices at $\widetilde{t}=0$, while panels (d)--(f) show the corresponding slices at $\widetilde{t}=2000$, after a compact Proca star has formed (in each panel, the density maximum has been shifted to the center).}
\label{fig:density-snapshots}
\end{figure}

Figure~\ref{fig:density-snapshots} shows the evolution of the density field for the three polarization preparations.  All simulations start from homogeneous and isotropic random fields.  After gravitational evolution, each realization develops a stable Proca star surrounded by an interference halo, consistent with our previous studies~\cite{PhysRevD.108.083021}.  With these dynamically formed Proca stars in place, we now focus on their spin properties.

\begin{figure}[t]
\centering
\includegraphics[width=0.98\linewidth]{\figpath{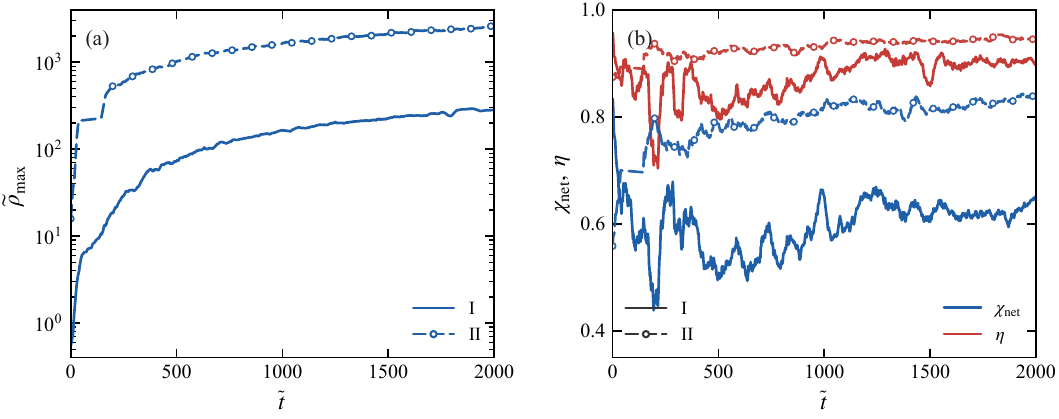}}
\caption{Proca-star formation and internal spin diagnostics.  (a) Maximum dimensionless density $\widetilde{\rho}_{\max}$ as a function of dimensionless time $\widetilde{t}$ for the two independent-component examples.  (b) Net spin fraction $\chi_{\rm net}(\widetilde{R}_c)$ and directional coherence $\eta(\widetilde{R}_c)=\chi_{\rm net}/\chi_{\rm loc}$ inside the half-maximum core aperture.  Blue denotes $\chi_{\rm net}$ and red denotes $\eta$; solid curves and dashed curves with open markers correspond to realizations I and II, respectively.}
\label{fig:prd-core-diagnostics}
\end{figure}

Figure~\ref{fig:prd-core-diagnostics} follows the density growth and spin diagnostics of two independent realizations.  In both cases the maximum density climbs quickly while the core assembles and then levels off.  The half-maximum aperture only becomes a meaningful core radius once that peak has settled, so the spin fractions before then trace the formation transient rather than the core.  The settled half-maximum aperture defines
\begin{equation}
\chi_{\rm net}(\widetilde{R}_c)=\frac{|\int_{\widetilde{r}<\widetilde{R}_c}\mathbf{s}\,d^3\widetilde{x}|}{\int_{\widetilde{r}<\widetilde{R}_c}\rho\,d^3\widetilde{x}}.
\end{equation}
For two independent-component examples, this quantity settles to sizable but submaximal values,
\begin{equation}
\chi_{\rm net}=0.621\pm0.062,\qquad 0.830\pm0.031.
\end{equation}
The corresponding local fractions are
\begin{equation}
\chi_{\rm loc}=0.688\pm0.060,\qquad 0.877\pm0.029.
\end{equation}
The corresponding coherence ratios are
\begin{equation}
\eta=0.902\pm0.043,\qquad 0.946\pm0.015.
\end{equation}
Most of the local spin density therefore points the same way inside the core aperture: the signal is a coherent internal polarization state, not a large $|\mathbf{s}|$ averaged over cancelling directions.

The higher-spin example matches the $\sim0.8$ core spin fraction found in the same condensation setting in Ref.~\cite{PhysRevD.108.083021}, and the decomposition clarifies the interpretation of that number: $\chi_{\rm net}$ is the coherent aperture-averaged spin, and $\eta\simeq1$ shows the directional cancellation inside the core is weak.  The pattern survives when the aperture is pulled inward (Table~\ref{tab:aperture-robustness}); $\widetilde{R}_c$ is the common core aperture below.

\begin{table}[t]
\centering
\small
\begin{tabular}{c|ccc|ccc}
\hline
 & \multicolumn{3}{c|}{realization I} & \multicolumn{3}{c}{realization II} \\
$\widetilde R/\widetilde R_c$ & $\chi_{\rm net}$ & $\chi_{\rm loc}$ & $\eta$ & $\chi_{\rm net}$ & $\chi_{\rm loc}$ & $\eta$ \\
\hline
$0.25$ & $0.707$ & $0.707$ & $1.000$ & $0.922$ & $0.922$ & $1.000$ \\
$0.50$ & $0.687$ & $0.702$ & $0.979$ & $0.922$ & $0.922$ & $1.000$ \\
$0.75$ & $0.661$ & $0.696$ & $0.950$ & $0.870$ & $0.897$ & $0.970$ \\
$1.00$ & $0.621$ & $0.688$ & $0.902$ & $0.830$ & $0.877$ & $0.946$ \\
\hline
\end{tabular}
\caption{Aperture robustness of the internal spin diagnostics for the two independent-component examples.  The entries are averages over the condensed late-time interval.  The identical first two entries for realization II occur because those two small apertures select the same central volume element.}
\label{tab:aperture-robustness}
\end{table}

Before a compact core has formed, $\widetilde R_c$ is only an instantaneous density-defined radius, so the early-time spin fractions in Fig.~\ref{fig:prd-core-diagnostics} are used as formation diagnostics rather than as part of the quoted late-time averages.

\subsection{Component Correlations and the Circular Limit}

Component correlations provide a controlled deformation of the core polarization.  The pointwise-circular preparation of Eq.~\eqref{eq:circular-xy-ic} stays at the bound, $\chi_{\rm net}(\widetilde{R}_c)=1$, and a partially correlated realization sits just inside it, with $\chi_{\rm net}(\widetilde{R}_c)=0.981$ and $\eta(\widetilde{R}_c)=0.997$.  A matched correlated--independent pair gives $\chi_{\rm net}(\widetilde{R}_c)=0.987\pm0.003$ and $0.653\pm0.038$, respectively.  The imposed correlation therefore drives the condensed core toward the circular-polarization limit, whereas independent components yield a submaximal polarization fraction.

These preparations are not intended as generic virialized halo initial conditions; persistent helicity bias would require special initial conditions or additional physics~\cite{Amaral:2024ezg}.

For independent components, we performed an extensive simulation suite spanning random seeds, component normalizations, and box sizes.  This suite gives
\begin{equation}
\langle\chi_{\rm net}(\widetilde{R}_c)\rangle=0.616,\qquad
\sigma_{\rm run}=0.159.
\label{eq:ensemble-spin}
\end{equation}
Here $\sigma_{\rm run}$ denotes the realization-to-realization scatter across the independent simulations.  The scatter reflects the fact that each realization selects a different component-space polarization.  Correlated preparations cluster near the circular limit, while the circular preparation remains at that limit by the preserved component relation in Eq.~\eqref{eq:circular-xy-ic}.

\subsection{Partial Spin from Random Complex Vectors}

The submaximal spin fractions follow once the core is approximated by a single dominant spatial mode,
\begin{equation}
\psi_i(\mathbf{x},t)\simeq c_i\phi(\mathbf{x},t)+\delta\psi_i ,
\label{eq:single-mode}
\end{equation}
where $\mathbf{c}=(c_x,c_y,c_z)$ is a complex vector and $\delta\psi_i$ collects residual excited modes, finite-aperture corrections, and imperfect component locking.  Dropping $\delta\psi_i$, the normalized core spin fraction reduces to
\begin{equation}
\chi_c=\frac{|i\mathbf{c}\times\mathbf{c}^{\ast}|}{\mathbf{c}^{\dagger}\mathbf{c}}.
\end{equation}
Thus $\chi_c$ is the ideal single-mode value approached by $\chi_{\rm net}$ when the residual $\delta\psi_i$ is negligible.  It equals unity only for a circular $\mathbf{c}$, and for independent-component initial data no symmetry pins $\mathbf{c}$ to the linear or circular case.  The condensate is then free to settle on a generic elliptical polarization, which gives an intermediate spin fraction --- $\chi_c=0$ for a linear vector, $\chi_c=1$ for a circular one.  A value near $0.8$ signals a strongly elliptical dominant mode, whereas correlated and circular initial data push the same fraction to the circular bound.

To make this statement quantitative, we compare with two simple single-mode distributions.  These distributions predict the ideal single-mode quantity $\chi_c$, not the finite-aperture $\chi_{\rm net}$ itself.  The comparison uses $\chi_{\rm eig}$ from the fixed-box, equal-component independent subset, with $\widetilde M_x=\widetilde M_y=\widetilde M_z$.  Its relation to the directly integrated $\chi_{\rm net}$ is tested separately in Fig.~\ref{fig:polarization-matrix}.  In the random-complex-vector model,
\begin{equation}
\mathbf c=\frac{\mathbf z}{(\mathbf z^\dagger\mathbf z)^{1/2}},\qquad
z_i=a_i+i b_i,\qquad a_i,b_i\sim{\cal N}(0,1),
\label{eq:random-complex-reference}
\end{equation}
with all six real Gaussian variables drawn independently.  This allows both the relative amplitudes and the relative phases of the three vector components to fluctuate.  To see the resulting distribution, write $\mathbf z=\mathbf a+i\mathbf b$.  Here $\sigma_1$ and $\sigma_2$ are the two singular values of the $3\times2$ real matrix $(\mathbf a,\mathbf b)$, representing the principal axes of the polarization ellipse.  They give $\chi_c=2\sigma_1\sigma_2/(\sigma_1^2+\sigma_2^2)$; integrating the corresponding $2\times2$ Wishart eigenvalue distribution yields
\begin{equation}
F_{\rm iso}(\chi_c)=\chi_c^2,\qquad
f_{\rm iso}(\chi_c)=2\chi_c,\qquad 0\leq\chi_c\leq1 .
\label{eq:random-complex-analytic}
\end{equation}
The simple form arises because the $3\times2$ real Gaussian matrix gives a $2\times2$ Wishart problem with three degrees of freedom; for other numbers of vector components the corresponding distribution would differ.  Equivalently, $\chi_c^2$ is uniformly distributed in this isotropic ensemble, giving $\langle\chi_c\rangle=2/3$ and $\chi_{\rm med}=1/\sqrt{2}$.

Motivated by the equal component numbers in this subset, we also consider a model in which the core retains equal component amplitudes while only the relative phases are randomized,
\begin{equation}
c_i=\frac{e^{i\theta_i}}{\sqrt{3}},\qquad
\theta_i\sim U(0,2\pi).
\label{eq:equal-amplitude-reference}
\end{equation}
Although the component amplitudes are fixed, $\chi_c$ remains a random variable because it depends on the relative phases $\theta_i$.  We therefore evaluate this distribution by Monte Carlo sampling, which gives a distribution shifted toward the circular bound, with mean $0.790$ and median $0.871$.  Its mismatch with the data indicates unequal core amplitudes, with the surrounding halo compensating this local redistribution.

\begin{figure}[!htbp]
\centering
\includegraphics[width=0.72\linewidth]{\figpath{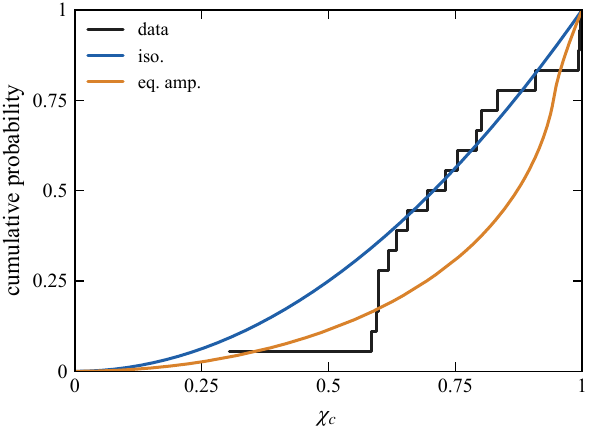}}
\caption{Cumulative distribution of the leading-eigenvector estimates $\chi_{\rm eig}$ for the fixed-box independent realizations, compared with the analytic random-complex-vector model in Eq.~\eqref{eq:random-complex-analytic} and the equal-amplitude random-phase model in Eq.~\eqref{eq:equal-amplitude-reference}.}
\label{fig:random-elliptical-reference}
\end{figure}

Figure~\ref{fig:random-elliptical-reference} compares these models with the measured leading-eigenvector estimates.  The empirical distribution follows the random-complex-vector model more closely than the equal-amplitude phase model, whose weight is concentrated too near the circular bound.  We use this comparison as a consistency check of the single-mode polarization interpretation, rather than as a precision fit to a universal distribution.

\subsection{Polarization Matrix and Rank-One Structure}

The core polarization matrix in Eq.~\eqref{eq:polarization-matrix} provides the direct link between the measured aperture spin and the component-space mode that carries it.  Equations~\eqref{eq:spin-from-p} and \eqref{eq:chi-from-p} show that its imaginary antisymmetric part is exactly $\mathbf{S}_c/N_c$ and $\chi_{\rm net}$, while Eqs.~\eqref{eq:p-eigenproblem}--\eqref{eq:chi-eig} define the eigenvalues, the purity, and the leading-eigenvector spin fraction.  Therefore a rank-one, single-mode core should have $\lambda_1\simeq1$, $\mathcal P\simeq1$, and $\chi_{\rm eig}\simeq\chi_{\rm net}$.
For the two independent examples, diagonalizing $P$ at the core aperture gives
\begin{equation}
(\lambda_1,\lambda_2,\lambda_3)=(0.915,0.058,0.027),\qquad
(0.925,0.051,0.024),
\end{equation}
with purities $\mathcal P=0.842$ and $0.859$: one direction carries more than $90\%$ of the weight.  The leading eigenvector gives $\chi_{\rm eig}=0.710$ and $0.927$, close to the directly integrated $0.621$ and $0.830$.  That the two nearly agree is the separate statement that the three components also share essentially one spatial profile, as assumed in Eq.~\eqref{eq:single-mode}; the small shortfall of the integrated spin comes from the subdominant modes and finite-aperture mixing left out of the leading vector.  Correlated and circular preparations sit near the circular bound for the same reason: their leading component-space vectors are themselves nearly circular.

\begin{figure}[!htbp]
\centering
\includegraphics[width=0.98\linewidth]{\figpath{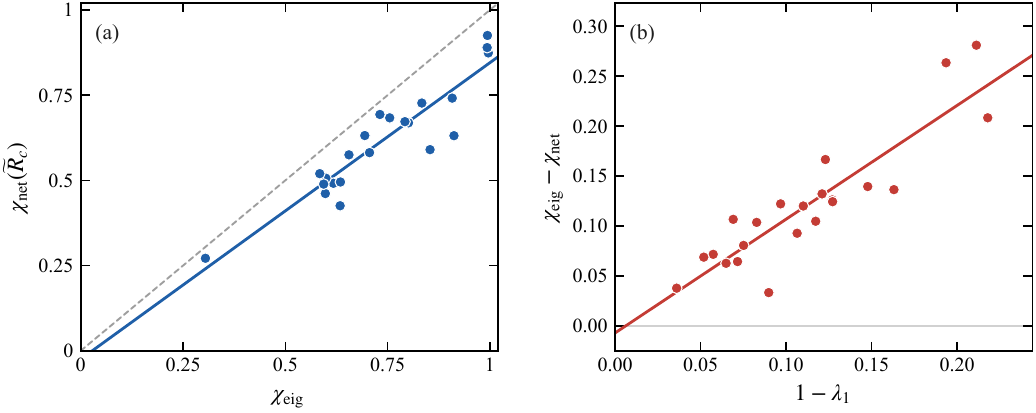}}
\caption{Polarization-matrix interpretation of the independent realizations.  (a) Directly integrated $\chi_{\rm net}(\widetilde{R}_c)$ versus the leading-eigenvector prediction $\chi_{\rm eig}$.  The gray line marks equality and the blue line shows $\chi_{\rm net}\simeq0.87\,\chi_{\rm eig}-0.024$.  (b) Difference $\chi_{\rm eig}-\chi_{\rm net}$ versus the rank-one residual $1-\lambda_1$.  The red line shows $\Delta\chi\simeq1.14(1-\lambda_1)-0.007$.}
\label{fig:polarization-matrix}
\end{figure}

Figure~\ref{fig:polarization-matrix}(a) plots the directly integrated $\chi_{\rm net}(\widetilde{R}_c)$ against $\chi_{\rm eig}$ across the independent realizations.  The two track each other closely,
\begin{equation}
\chi_{\rm net}(\widetilde{R}_c)\simeq(0.87\pm0.08)\,\chi_{\rm eig}-(0.024\pm0.060) ,
\end{equation}
with Pearson $r\simeq0.93$, and $\chi_{\rm eig}$ sits systematically above $\chi_{\rm net}$ because it keeps only the dominant mode while the aperture integral also samples the departures from a rank-one core.  Those departures set the gap: writing $\Delta\chi=\chi_{\rm eig}-\chi_{\rm net}(\widetilde{R}_c)$, Fig.~\ref{fig:polarization-matrix}(b) gives
\begin{equation}
\Delta\chi\simeq(1.14\pm0.13)(1-\lambda_1)-(0.007\pm0.015) ,
\end{equation}
with $r\simeq0.90$ and $R^2\simeq0.80$.  The intercept is consistent with zero, so $\Delta\chi\to0$ as $\lambda_1\to1$: in the exact single-mode limit the eigenvector prediction and the aperture integral coincide, as they must.

These results support the following interpretation.  The common gravitational potential collects the three components into the same density peak, while the internal spin is set by the amplitudes and phases of the dominant component-space mode.  The box-integrated component-space matrix is conserved, while the core matrix can evolve through exchange with the surrounding halo.  A Proca star can therefore acquire a large local polarization even when the box-integrated spin is small, with the surrounding interference halo carrying the compensating component-space polarization.  Independent components land on an elliptical mode and a partial spin fraction; correlated or circular data land on a nearly circular one, where the average spin density approaches $\rho$ and $\chi\simeq1$.  The ordering independent $\rightarrow$ correlated $\rightarrow$ circular is the expected ordering of component-space polarization coherence.

\section{Conclusion}
\label{sec:conclusion}

We have developed a quantitative framework for the internal spin polarization of Proca stars formed by gravitational Bose--Einstein condensation in the nonrelativistic vector-soliton regime.  Its central step splits the aperture spin into a coherent net fraction $\chi_{\rm net}$, a local fraction $\chi_{\rm loc}$, and the ratio $\eta=\chi_{\rm net}/\chi_{\rm loc}$, which distinguishes a genuinely coherent core spin from local polarization that cancels over directions.

Independent-component condensation leaves the core sizably but not maximally polarized, with $\eta$ close to unity.  The ensemble mean and scatter, $\langle\chi_{\rm net}(\widetilde R_c)\rangle=0.616$ and $\sigma_{\rm run}=0.159$, place the $\sim0.8$ values reported earlier~\cite{PhysRevD.108.083021} in context: they are high-spin realizations of a random-polarization outcome rather than evidence for a universal Proca-star spin fraction.

This distribution is naturally interpreted in terms of the component-space polarization of the dominant mode.  In the single-mode limit $\psi_i\simeq c_i\phi$, the spin fraction is set by the elliptical polarization of the complex vector $\mathbf c$: independent components select generic elliptical vectors, whereas correlated and circular preparations select nearly circular vectors that approach the bound $\chi=1$.  The core polarization matrix supports this single-mode interpretation.  Its leading-eigenvector spin fraction tracks the directly integrated core spin, and the residual is controlled by the departure from a rank-one core.

Internal polarization is therefore an intrinsic degree of freedom of vector condensates.  A scalar solitonic core is characterized mainly by its density and phase, whereas a nonrelativistic Proca star also carries a component-space polarization state built during condensation.  The present simulations are idealized and non-cosmological, but the same diagnostics can be used to analyze condensed Proca stars in cosmological vector-dark-matter simulations.  It will be important to determine whether the polarization distribution found here is inherited, erased, or regenerated during hierarchical assembly and core mergers.  If such polarization survives in realistic environments, it would provide a vector-specific imprint on small-scale structure beyond the scalar density profile.

\acknowledgments
The author thanks Zhipan Li for computational assistance and Jiajie Li for helpful comments on the manuscript.  The author also thanks Mudit Jain for helpful comments and suggestions that improved the manuscript.  J.C. acknowledges support from the Fundamental Research Funds for the Central Universities under Grant No.~SWU-KR22012 and from the Chongqing Natural Science Foundation General Project under Grant No.~CSTB2023NSCQ-MSX0453.  This work was also supported by computational resources provided by the School of Physical Science and Technology at Southwest University.

\bibliographystyle{jhep}
\bibliography{references}

\providecommand{\href}[2]{#2}\begingroup\raggedright\begin{thebibliography}{10}

\bibitem{Aghanim:2018eyx}
{\scshape Planck} collaboration, \emph{Planck 2018 results. vi. cosmological
  parameters}, \href{https://doi.org/10.1051/0004-6361/201833910}{\emph{Astron.
  Astrophys.} {\bfseries 641} (2020) A6}
  [\href{https://arxiv.org/abs/1807.06209}{{\ttfamily 1807.06209}}].

\bibitem{hu2000}
W.~Hu, R.~Barkana and A.~Gruzinov, \emph{Cold and fuzzy dark matter},
  \href{https://doi.org/10.1103/PhysRevLett.85.1158}{\emph{Phys. Rev. Lett.}
  {\bfseries 85} (2000) 1158}
  [\href{https://arxiv.org/abs/astro-ph/0003365}{{\ttfamily
  astro-ph/0003365}}].

\bibitem{Sikivie:2006ni}
P.~Sikivie, \emph{Axion cosmology},
  \href{https://doi.org/10.1007/978-3-540-73518-2_2}{\emph{Lect. Notes Phys.}
  {\bfseries 741} (2008) 19}
  [\href{https://arxiv.org/abs/astro-ph/0610440}{{\ttfamily
  astro-ph/0610440}}].

\bibitem{Arvanitaki:2009fg}
A.~Arvanitaki, S.~Dimopoulos, S.~Dubovsky, N.~Kaloper and J.~March-Russell,
  \emph{String axiverse},
  \href{https://doi.org/10.1103/PhysRevD.81.123530}{\emph{Phys. Rev. D}
  {\bfseries 81} (2010) 123530}
  [\href{https://arxiv.org/abs/0905.4720}{{\ttfamily 0905.4720}}].

\bibitem{Hui:2016ltb}
L.~Hui, J.~P. Ostriker, S.~Tremaine and E.~Witten, \emph{Ultralight scalars as
  cosmological dark matter},
  \href{https://doi.org/10.1103/PhysRevD.95.043541}{\emph{Phys. Rev. D}
  {\bfseries 95} (2017) 043541}
  [\href{https://arxiv.org/abs/1610.08297}{{\ttfamily 1610.08297}}].

\bibitem{Marsh:2015xka}
D.~J.~E. Marsh, \emph{Axion cosmology},
  \href{https://doi.org/10.1016/j.physrep.2016.06.005}{\emph{Phys. Rept.}
  {\bfseries 643} (2016) 1} [\href{https://arxiv.org/abs/1510.07633}{{\ttfamily
  1510.07633}}].

\bibitem{Suarez:2013iw}
A.~Suarez, V.~H. Robles and T.~Matos, \emph{A review on the scalar
  field/bose-einstein condensate dark matter model},
  \href{https://doi.org/10.1007/978-3-319-02063-1_9}{\emph{Astrophys. Space
  Sci. Proc.} {\bfseries 38} (2014) 107}
  [\href{https://arxiv.org/abs/1302.0903}{{\ttfamily 1302.0903}}].

\bibitem{Hlozek:2017zzf}
R.~Hlozek, D.~J.~E. Marsh and D.~Grin, \emph{Using the full power of the cosmic
  microwave background to probe axion dark matter},
  \href{https://doi.org/10.1093/mnras/sty271}{\emph{Mon. Not. Roy. Astron.
  Soc.} {\bfseries 476} (2018) 3063}
  [\href{https://arxiv.org/abs/1708.05681}{{\ttfamily 1708.05681}}].

\bibitem{Schive:2014dra}
H.-Y. Schive, T.~Chiueh and T.~Broadhurst, \emph{Cosmic structure as the
  quantum interference of a coherent dark wave},
  \href{https://doi.org/10.1038/nphys2996}{\emph{Nature Phys.} {\bfseries 10}
  (2014) 496} [\href{https://arxiv.org/abs/1406.6586}{{\ttfamily 1406.6586}}].

\bibitem{Schive:2014hza}
H.-Y. Schive, M.-H. Liao, T.~Woo, S.-K. Wong, T.~Chiueh, T.~Broadhurst et~al.,
  \emph{Understanding the core-halo relation of quantum wave dark matter from
  3d simulations},
  \href{https://doi.org/10.1103/PhysRevLett.113.261302}{\emph{Phys. Rev. Lett.}
  {\bfseries 113} (2014) 261302}
  [\href{https://arxiv.org/abs/1407.7762}{{\ttfamily 1407.7762}}].

\bibitem{Veltmaat:2018dfz}
J.~Veltmaat, J.~C. Niemeyer and B.~Schwabe, \emph{Formation and structure of
  ultralight bosonic dark matter halos},
  \href{https://doi.org/10.1103/PhysRevD.98.043509}{\emph{Phys. Rev. D}
  {\bfseries 98} (2018) 043509}
  [\href{https://arxiv.org/abs/1804.09647}{{\ttfamily 1804.09647}}].

\bibitem{Bar:2018acw}
N.~Bar, D.~Blas, K.~Blum and S.~Sibiryakov, \emph{Galactic rotation curves
  versus ultralight dark matter: Implications of the soliton-host halo
  relation}, \href{https://doi.org/10.1103/PhysRevD.98.083027}{\emph{Phys. Rev.
  D} {\bfseries 98} (2018) 083027}
  [\href{https://arxiv.org/abs/1805.00122}{{\ttfamily 1805.00122}}].

\bibitem{Du:2016aik}
X.~Du, C.~Behrens, J.~C. Niemeyer and B.~Schwabe, \emph{Core-halo mass relation
  of ultralight axion dark matter from merger history},
  \href{https://doi.org/10.1103/PhysRevD.95.043519}{\emph{Phys. Rev. D}
  {\bfseries 95} (2017) 043519}
  [\href{https://arxiv.org/abs/1609.09414}{{\ttfamily 1609.09414}}].

\bibitem{Mocz:2017wlg}
P.~Mocz, M.~Vogelsberger, V.~Robles, J.~Zavala, M.~Boylan-Kolchin and
  L.~Hernquist, \emph{Galaxy formation with becdm: I. turbulence and relaxation
  of idealised haloes}, \href{https://doi.org/10.1093/mnras/stx1887}{\emph{Mon.
  Not. Roy. Astron. Soc.} {\bfseries 471} (2017) 4559}
  [\href{https://arxiv.org/abs/1705.05845}{{\ttfamily 1705.05845}}].

\bibitem{Graham:2015rva}
P.~W. Graham, J.~Mardon and S.~Rajendran, \emph{Vector dark matter from
  inflationary fluctuations},
  \href{https://doi.org/10.1103/PhysRevD.93.103520}{\emph{Phys. Rev. D}
  {\bfseries 93} (2016) 103520}
  [\href{https://arxiv.org/abs/1504.02102}{{\ttfamily 1504.02102}}].

\bibitem{PhysRevD.108.083021}
J.~Chen, X.~Du, M.~Zhou, A.~Benson and D.~J.~E. Marsh, \emph{Gravitational
  bose-einstein condensation of vector or hidden photon dark matter},
  \href{https://doi.org/10.1103/PhysRevD.108.083021}{\emph{Phys. Rev. D}
  {\bfseries 108} (2023) 083021}.

\bibitem{PhysRevD.111.043031}
J.~Chen, L.~H. Nguyen and D.~J.~E. Marsh, \emph{Vector dark matter halo: From
  polarization dynamics to direct detection},
  \href{https://doi.org/10.1103/PhysRevD.111.043031}{\emph{Phys. Rev. D}
  {\bfseries 111} (2025) 043031}.

\bibitem{Jain:2021pnk}
M.~Jain and M.~A. Amin, \emph{Polarized solitons in higher-spin wave dark
  matter}, \href{https://doi.org/10.1103/PhysRevD.105.056019}{\emph{Phys. Rev.
  D} {\bfseries 105} (2022) 056019}
  [\href{https://arxiv.org/abs/2109.04892}{{\ttfamily 2109.04892}}].

\bibitem{Amin:2022pkn}
M.~A. Amin, M.~Jain, R.~Karur and P.~Mocz, \emph{Small-scale structure in
  vector dark matter},
  \href{https://doi.org/10.1088/1475-7516/2022/08/014}{\emph{JCAP} {\bfseries
  2022} (2022) 014} [\href{https://arxiv.org/abs/2203.11935}{{\ttfamily
  2203.11935}}].

\bibitem{Gorghetto:2022sue}
M.~Gorghetto, E.~Hardy, J.~March-Russell, N.~Song and S.~M. West, \emph{Dark
  photon stars: formation and role as dark matter substructure},
  \href{https://arxiv.org/abs/2203.10100}{{\ttfamily 2203.10100}}.

\bibitem{Volkov:2002aj}
M.~S. Volkov and E.~Wohnert, \emph{Spinning q-balls},
  \href{https://doi.org/10.1103/PhysRevD.66.085003}{\emph{Phys. Rev. D}
  {\bfseries 66} (2002) 085003}
  [\href{https://arxiv.org/abs/hep-th/0205157}{{\ttfamily hep-th/0205157}}].

\bibitem{Zhang:2024bjo}
H.-Y. Zhang, \emph{Unified view of scalar and vector dark matter solitons},
  \href{https://doi.org/10.1007/JHEP04(2025)174}{\emph{JHEP} {\bfseries 04}
  (2025) 174} [\href{https://arxiv.org/abs/2406.05031}{{\ttfamily
  2406.05031}}].

\bibitem{1991PhRvL..66.1659S}
E.~Seidel and W.-M. Suen, \emph{Oscillating soliton stars},
  \href{https://doi.org/10.1103/PhysRevLett.66.1659}{\emph{Phys. Rev. Lett.}
  {\bfseries 66} (1991) 1659}.

\bibitem{PhysRevD.42.384}
E.~Seidel and W.-M. Suen, \emph{Dynamical evolution of boson stars: Perturbing
  the ground state}, \href{https://doi.org/10.1103/PhysRevD.42.384}{\emph{Phys.
  Rev. D} {\bfseries 42} (1990) 384}.

\bibitem{1994PhRvL..72.2516S}
E.~Seidel and W.-M. Suen, \emph{Formation of solitonic stars through
  gravitational cooling},
  \href{https://doi.org/10.1103/PhysRevLett.72.2516}{\emph{Phys. Rev. Lett.}
  {\bfseries 72} (1994) 2516}
  [\href{https://arxiv.org/abs/gr-qc/9309015}{{\ttfamily gr-qc/9309015}}].

\bibitem{Liddle:1993ha}
A.~R. Liddle and M.~S. Madsen, \emph{The structure and formation of boson
  stars}, \href{https://doi.org/10.1142/S0218271892000057}{\emph{Int. J. Mod.
  Phys. D} {\bfseries 1} (1992) 101}.

\bibitem{Kolb:1993zz}
E.~W. Kolb and I.~I. Tkachev, \emph{Axion miniclusters and bose stars},
  \href{https://doi.org/10.1103/PhysRevLett.71.3051}{\emph{Phys. Rev. Lett.}
  {\bfseries 71} (1993) 3051}
  [\href{https://arxiv.org/abs/hep-ph/9303313}{{\ttfamily hep-ph/9303313}}].

\bibitem{Kolb:1993hw}
E.~W. Kolb and I.~I. Tkachev, \emph{Nonlinear axion dynamics and formation of
  cosmological pseudosolitons},
  \href{https://doi.org/10.1103/PhysRevD.49.5040}{\emph{Phys. Rev. D}
  {\bfseries 49} (1994) 5040}
  [\href{https://arxiv.org/abs/astro-ph/9311037}{{\ttfamily
  astro-ph/9311037}}].

\bibitem{PhysRevD.84.043531}
P.-H. Chavanis, \emph{Mass-radius relation of newtonian self-gravitating
  bose-einstein condensates with short-range interactions. i. analytical
  results}, \href{https://doi.org/10.1103/PhysRevD.84.043531}{\emph{Phys. Rev.
  D} {\bfseries 84} (2011) 043531}.

\bibitem{Chavanis:2011zm}
P.-H. Chavanis and L.~Delfini, \emph{Mass-radius relation of newtonian
  self-gravitating bose-einstein condensates with short-range interactions: Ii.
  numerical results},
  \href{https://doi.org/10.1103/PhysRevD.84.043532}{\emph{Phys. Rev. D}
  {\bfseries 84} (2011) 043532}
  [\href{https://arxiv.org/abs/1103.2054}{{\ttfamily 1103.2054}}].

\bibitem{Chavanis:2016dab}
P.-H. Chavanis, \emph{Collapse of a self-gravitating bose-einstein condensate
  with attractive self-interaction},
  \href{https://doi.org/10.1103/PhysRevD.94.083007}{\emph{Phys. Rev. D}
  {\bfseries 94} (2016) 083007}
  [\href{https://arxiv.org/abs/1604.05904}{{\ttfamily 1604.05904}}].

\bibitem{Eby:2015hsq}
J.~Eby, C.~Kouvaris, N.~G. Nielsen and L.~C.~R. Wijewardhana, \emph{Boson stars
  from self-interacting dark matter},
  \href{https://doi.org/10.1007/JHEP02(2016)028}{\emph{JHEP} {\bfseries 2016}
  (2016) 028} [\href{https://arxiv.org/abs/1511.04474}{{\ttfamily
  1511.04474}}].

\bibitem{Levkov:2018kau}
D.~G. Levkov, A.~G. Panin and I.~I. Tkachev, \emph{Gravitational bose-einstein
  condensation in the kinetic regime},
  \href{https://doi.org/10.1103/PhysRevLett.121.151301}{\emph{Phys. Rev. Lett.}
  {\bfseries 121} (2018) 151301}
  [\href{https://arxiv.org/abs/1804.05857}{{\ttfamily 1804.05857}}].

\bibitem{PhysRevD.104.083022}
J.~Chen, X.~Du, E.~W. Lentz, D.~J.~E. Marsh and J.~C. Niemeyer, \emph{New
  insights into the formation and growth of boson stars in dark matter halos},
  \href{https://doi.org/10.1103/PhysRevD.104.083022}{\emph{Phys. Rev. D}
  {\bfseries 104} (2021) 083022}.

\bibitem{PhysRevD.106.023009}
J.~Chen, X.~Du, E.~W. Lentz and D.~J.~E. Marsh, \emph{Relaxation times for
  bose-einstein condensation by self-interaction and gravity},
  \href{https://doi.org/10.1103/PhysRevD.106.023009}{\emph{Phys. Rev. D}
  {\bfseries 106} (2022) 023009}.

\bibitem{Amin:2019ums}
M.~A. Amin and P.~Mocz, \emph{Formation, gravitational clustering, and
  interactions of nonrelativistic solitons in an expanding universe},
  \href{https://doi.org/10.1103/PhysRevD.100.063507}{\emph{Phys. Rev. D}
  {\bfseries 100} (2019) 063507}
  [\href{https://arxiv.org/abs/1902.07261}{{\ttfamily 1902.07261}}].

\bibitem{Kirkpatrick:2020fwd}
K.~Kirkpatrick, A.~E. Mirasola and C.~Prescod-Weinstein, \emph{Relaxation times
  for bose-einstein condensation in axion miniclusters},
  \href{https://doi.org/10.1103/PhysRevD.102.103012}{\emph{Phys. Rev. D}
  {\bfseries 102} (2020) 103012}
  [\href{https://arxiv.org/abs/2007.07438}{{\ttfamily 2007.07438}}].

\bibitem{Eggemeier:2019jsu}
B.~Eggemeier and J.~C. Niemeyer, \emph{Formation and mass growth of axion stars
  in axion miniclusters},
  \href{https://doi.org/10.1103/PhysRevD.100.063528}{\emph{Phys. Rev. D}
  {\bfseries 100} (2019) 063528}
  [\href{https://arxiv.org/abs/1906.01348}{{\ttfamily 1906.01348}}].

\bibitem{Jain:2023ojg}
M.~Jain, M.~A. Amin, J.~Thomas and W.~Wanichwecharungruang, \emph{Kinetic
  relaxation and bose-star formation in multicomponent dark matter},
  \href{https://doi.org/10.1103/PhysRevD.108.043535}{\emph{Phys. Rev. D}
  {\bfseries 108} (2023) 043535}
  [\href{https://arxiv.org/abs/2304.01985}{{\ttfamily 2304.01985}}].

\bibitem{Jain:2024krs}
M.~Jain, W.~Wanichwecharungruang and J.~Thomas, \emph{Kinetic relaxation and
  nucleation of bose stars in self-interacting wave dark matter},
  \href{https://doi.org/10.1103/PhysRevD.109.016002}{\emph{Phys. Rev. D}
  {\bfseries 109} (2024) 016002}
  [\href{https://arxiv.org/abs/2310.00058}{{\ttfamily 2310.00058}}].

\bibitem{Widrow:1993qq}
L.~M. Widrow and N.~Kaiser, \emph{Using the schroedinger equation to simulate
  collisionless matter}, \href{https://doi.org/10.1086/187073}{\emph{Astrophys.
  J. Lett.} {\bfseries 416} (1993) L71}.

\bibitem{Schwabe_2016}
B.~Schwabe, J.~C. Niemeyer and J.~F. Engels, \emph{Simulations of solitonic
  core mergers in ultralight axion dark matter cosmologies},
  \href{https://doi.org/10.1103/PhysRevD.94.043513}{\emph{Phys. Rev. D}
  {\bfseries 94} (2016) 043513}.

\bibitem{Mocz:2015sda}
P.~Mocz and S.~Succi, \emph{Numerical solution of the non-linear schr\"odinger
  equation using smoothed-particle hydrodynamics},
  \href{https://doi.org/10.1103/PhysRevE.91.053304}{\emph{Phys. Rev. E}
  {\bfseries 91} (2015) 053304}
  [\href{https://arxiv.org/abs/1503.03869}{{\ttfamily 1503.03869}}].

\bibitem{Uhlemann:2014npa}
C.~Uhlemann, M.~Kopp and T.~Haugg, \emph{Schr\"odinger method as $n$-body
  double and uv completion of dust},
  \href{https://doi.org/10.1103/PhysRevD.90.023517}{\emph{Phys. Rev. D}
  {\bfseries 90} (2014) 023517}
  [\href{https://arxiv.org/abs/1403.5567}{{\ttfamily 1403.5567}}].

\bibitem{Hui_2021}
L.~Hui, A.~Joyce, M.~J. Landry and X.~Li, \emph{Vortices and waves in light
  dark matter},
  \href{https://doi.org/10.1088/1475-7516/2021/01/011}{\emph{JCAP} {\bfseries
  2021} (2021) 011}.

\bibitem{Helfer_2019}
T.~Helfer, E.~A. Lim, M.~A.~G. Garcia and M.~A. Amin, \emph{Gravitational wave
  emission from collisions of compact scalar solitons},
  \href{https://doi.org/10.1103/PhysRevD.99.044046}{\emph{Phys. Rev. D}
  {\bfseries 99} (2019) 044046}.

\bibitem{Chen_2024}
J.~Chen and H.-Y. Zhang, \emph{Novel structures and collapse of solitons in
  nonminimally gravitating dark matter halos},
  \href{https://doi.org/10.1088/1475-7516/2024/10/005}{\emph{JCAP} {\bfseries
  2024} (2024) 005}.

\bibitem{Zeng:2026vortex}
Y.~Zeng, B.~Zhang and J.~Chen, \emph{Self-interaction controls vortex scale in
  soliton mergers}, \href{https://doi.org/10.1103/sm3l-z7s3}{\emph{Phys. Rev.
  D} {\bfseries 113} (2026) 103043}.

\bibitem{Brito:2015pxa}
R.~Brito, V.~Cardoso, C.~A.~R. Herdeiro and E.~Radu, \emph{Proca stars:
  Gravitating bose-einstein condensates of massive spin 1 particles},
  \href{https://doi.org/10.1016/j.physletb.2015.11.051}{\emph{Phys. Lett. B}
  {\bfseries 752} (2016) 291}
  [\href{https://arxiv.org/abs/1508.05395}{{\ttfamily 1508.05395}}].

\bibitem{Mendonca:2021aeq}
J.~T. Mendonca, \emph{Schr\"odinger-newton model with a background},
  \href{https://doi.org/10.3390/sym13061007}{\emph{Symmetry} {\bfseries 13}
  (2021) 1007}.

\bibitem{Salehian:2021khb}
B.~Salehian, H.-Y. Zhang, M.~A. Amin, D.~I. Kaiser and M.~H. Namjoo,
  \emph{Beyond schr\"odinger-poisson: nonrelativistic effective field theory
  for scalar dark matter},
  \href{https://doi.org/10.1007/JHEP09(2021)050}{\emph{JHEP} {\bfseries 2021}
  (2021) 050} [\href{https://arxiv.org/abs/2104.10128}{{\ttfamily
  2104.10128}}].

\bibitem{Amaral:2024ezg}
D.~W.~P. Amaral, M.~Jain, M.~A. Amin and C.~Tunnell, \emph{Vector wave dark
  matter and terrestrial quantum sensors},
  \href{https://doi.org/10.1088/1475-7516/2024/06/050}{\emph{JCAP} {\bfseries
  2024} (2024) 050} [\href{https://arxiv.org/abs/2403.02381}{{\ttfamily
  2403.02381}}].

\end{thebibliography}\endgroup

\end{document}